# Spin wave dynamics in artificial anti spin-ice systems: experimental and theoretical investigations

S. Mamica,[1] X. Zhou,[2] A. Adeyeye,[2] M. Krawczyk,[1] and G. Gubbiotti[3]

[1]Faculty of Physics, Adam Mickiewicz University in Poznan, Umultowska 85, 61-614 Poznan, Poland

[2]Information Storage Materials Laboratory, Department of Electrical and Computer Engineering, National University of Singapore, 117576 Singapore.

[3]Istituto Officina dei Materiali del CNR (CNR-IOM), Sede Secondaria di Perugia, c/o Dipartimento di Fisica e Geologia, Università di Perugia, I-06123 Perugia, Italy

**Abstract**

Reversed structures of artificial spin-ice systems, where elongated holes with elliptical shape (antidots) are arranged into a square array with two orthogonal sublattices, are referred to as anti-squared spin-ice. Using Brillouin light scattering spectroscopy and plane wave method calculations, we investigate the spin wave propagation perpendicular to the applied field direction for two 20 nm thick Permalloy nanostructures which differ by the presence of single and double elliptical antidots. For the spin waves propagation along the principal antidot lattice axis, the spectrum consists of flat bands separated by several frequency gaps which are the effect of spin wave amplitude confinement in the regions between antidots. Contrarily, for propagation direction at 45 degrees with respect to the antidot symmetry axis, straight and narrow channels of propagation are formed, leading to broadening of bands and closing of the magnonics gaps. Interestingly, in this case, extra magnonic band gaps occur due to the additional periodicity along this direction. The width and the position of these gaps depend on the presence of single or double antidots. In this context, we discuss possibilities for the tuning of spin wave spectra in anti-squared spin ice structures.

1. **Introduction**

Artificial spin-ice systems, i.e. close-packed arrays of single domain magnetic particles, have been largely investigated in the recent past.[1,2,3,4,5] The magnetic elements in these structures are made of soft ferromagnetic materials and are shaped with sizes too small to support more than one magnetic domain while small edge-to-edge distance allows significant magnetic stray fields coupling between neighboring elements that can extend across several neighbors. Usually, particles have the shape of elongated rectangular or elliptical dots with one long axis, such that a large uniaxial shape anisotropy dominates the magnetic properties, and geometric arrangements are chosen to introduce magnetic frustration in analogy to microscopic spin-ice studied in the bulk materials.[6,7] The spin wave (SW) properties in the GHz frequency range have been investigated in individual rectangular magnetic islands, in clusters and in arrays [8,9,10,11,12] as a function of magnetic field strength and orientation. For example, several groups have studied eigenmodes of a two sub-lattice elliptic-disks array aligned parallel and orthogonal to an applied magnetic field.[13,14,15]

In turn, the magnetization dynamics of the corresponding reverse structures, defined as anti-spin ice (ASI), have been scarcely investigated.[16,17] The ASI system is similar to that of an antidots lattice (ADL), i.e., there is a lattice of holes (antidots) drilled in a thin film.[18,19,20,21,22] The principal difference is determined by the rectangular or elliptical holes shape accompanied by their specific arrangement. As a consequence, the film is divided into regions which for different directions of the SW propagation and can be treated as isolated or connected.[23] This leads to specific effects in the SW dispersion.

In this paper, we provide the analysis of the SW spectra in two anti-square spin-ice lattices based on single, anti-square spin ice (ASSI), and double, anti-coupled square spin ice (ACSSI), antidots in the shape of elliptical holes. We studied the SW dispersion measured by Brillouin light scattering (BLS) spectroscopy for two different in-plane directions of the magnetic field applied along the square lattice axis ($\phi = 0°$) and along its diagonal ($\phi = 45°$). The obtained results are theoretically explained by using the plane wave method (PWM), which has proven to be a very efficient tool for calculating both the frequencies and the spatial profiles of excitations in different types of periodic structures such as photonic,[24] phononic,[25,26] magnonic[27] or magphonic (magneto-phononic) crystals.[28] In the case of magnonic crystals (MCs), this approach is useful for the periodicity of any dimensionality from 1D to 3D, as well as for quasiperiodic structures.[29,30,31,32,33]

We show that the specific structure of the system leads to the channeling of the SW propagation in certain directions. This is a consequence of the SW amplitude concentration in regions separated by antidots or in channels between antidots depending on the wave vector direction. In the first case SW confinement leads to the non-propagating waves reflected in flat bands in the dispersion relation while in the second case SWs freely propagate in channels. The specific structure of the investigated systems and the double periodicity existing in some particular directions, cause additional band gaps to open. Comparing results for ASSI and ACSSI bring us to some conclusions concerning possibilities for the tuning of the SW spectrum.

The paper is organized as follows. In the following three sections we describe the sample fabrication process, the BLS measurements details and the theoretical approach used, respectively. In the fifth section, we discuss the obtained results both from the experimental and theoretical point of view. In section six, we provide theoretical analysis concerning possibilities for the tuning of SW spectra. The paper is completed with conclusions.

**2. Sample fabrication**

The array of two-dimensional nano-islands over an area of 4×4 mm$^2$ was patterned on a resist film using deep ultraviolet lithography at 193 nm exposure wavelength. Details of the fabrication process for the 248 nm exposure wavelength system, which has been adapted for this work, have been described elsewhere.[34] Shown in Fig. 1(a) is a sketch of the fabrication process for the ASI structures. A 60 nm thick $Al_2O_3$ film was deposited on patterned resist template with electron beam deposition, followed by the ultrasonic-assisted lift-off process and bottom antireflection coatings removal process using ozone stripper. This was followed by the deposition of 20 nm thick Permalloy ($Ni_{80}Fe_{20}$, Py) film at a rate of 0.2 Å/s with a base pressure of $4 \times 10^{-8}$ Torr on top of the patterned $Al_2O_3$ nano-islands. The last process step is the removal of the $Al_2O_3$ which was achieved using AZ MIF 300 developer. Scanning electron microscopy (SEM) was used to verify the completion of lift-off and to confirm the critical dimensions of the pattern. Shown in Fig. 1 (b) is the representative SEM micrograph of the ASSI structures taken after the removal of $Al_2O_3$ patterns. As labeled in the image, the dimensions of the holes are 460 nm × 230 nm. The spacing between the nearest neighboring holes is 240 nm. By introducing a dimer, another structure, as shown in Fig. 1(c), with coupled holes was also fabricated. The gap between adjacent holes in ACSSI is around 60 nm. The lattice constant of the 2D square lattice is 900 nm and 1200 nm for the ASSI and ACSSI samples, respectively.

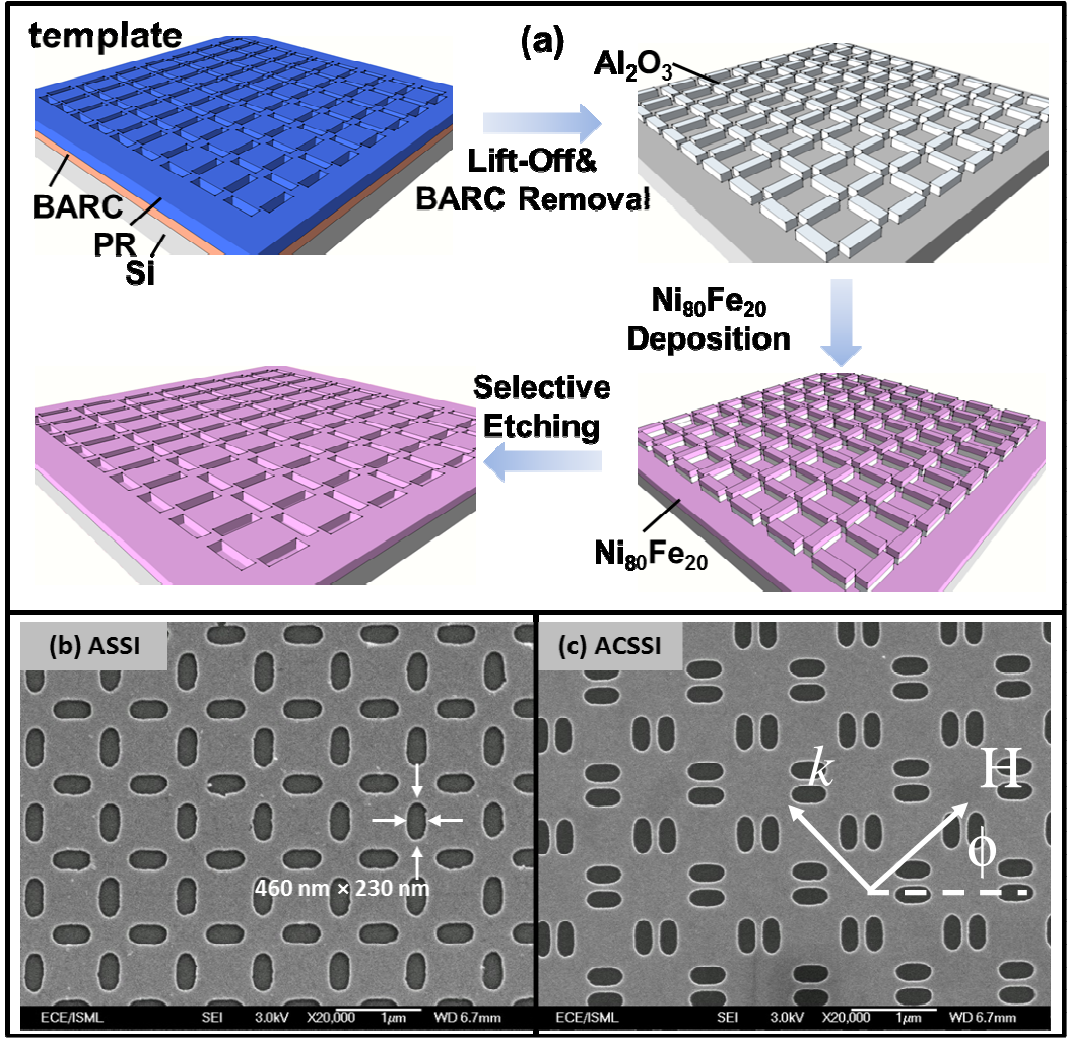

**Fig. 1.** (a) Schematic diagrams showing detailed fabrication processes to get the reverse structures. Scanning electron microscopy images of (b) anti-square spin ice (ASSI) and (c) anti-coupled square spin ice (ACSSI), together with the field direction and the scattering geometry. Here ϕ is the angle between the applied field (*H*) and the principal antidot lattice axis.

## 2. BLS measurements

BLS spectra from thermally excited SWs were recorded in the backscattering configuration where the incident light, from a single-mode solid-state laser operating at λ = 532 nm, is focused on the sample by a camera objective of numerical aperture NA = 0.24.[35] The scattered light is collected by the same objective and is then sent to the Sandercock-type (3+3)-tandem Fabry-Pérot interferometer to be analyzed in frequency. This geometry allows one to maximize the in-plane component of the wave vector exchanged in the scattering process and to select the wave vector (*k*) through the choice of the incidence angle (θ), $k = (4\pi/\lambda) \times \sin(\theta)$, in order to map the SW dispersion relation (frequency vs wave vector). Spectra were collected for different in-plane directions of the external magnetic field ($\phi = 0°$ and $\phi = 45°$), whose magnitude is fixed at $H = 0.14$ T [see Fig. 1(c) for the definition of $\phi$]. The direction of $H$ is measured with respect to the horizontal direction, along the chains of holes, and all the spectra were collected with $k$ perpendicular to $H$ [Damon-Eshbach (DE) configuration]. This field value ensures the sample saturation, as can be inferred from inspection of the measured hysteresis loops presented in Ref. 34. Fig. 2 shows the comparison between BLS spectra measured at normal light incidence ($k = 0$) for the ASSI and ACSSI samples for the two investigated field orientations, i.e. $\phi = 0°$ and $\phi = 45°$. Spectra measured for $\phi = 0°$ are quite similar for the two samples, showing a narrow very intense peak and another with smaller intensity at the lower frequency. The number of peaks increases when the field rotates by 45°. These features are further investigated in dependence on the wave vector magnitude, the results are shown in Fig. 3 and Fig. 5, where the measured dispersion for the ASSI and ACSSI, respectively, is compared to PWM calculations.

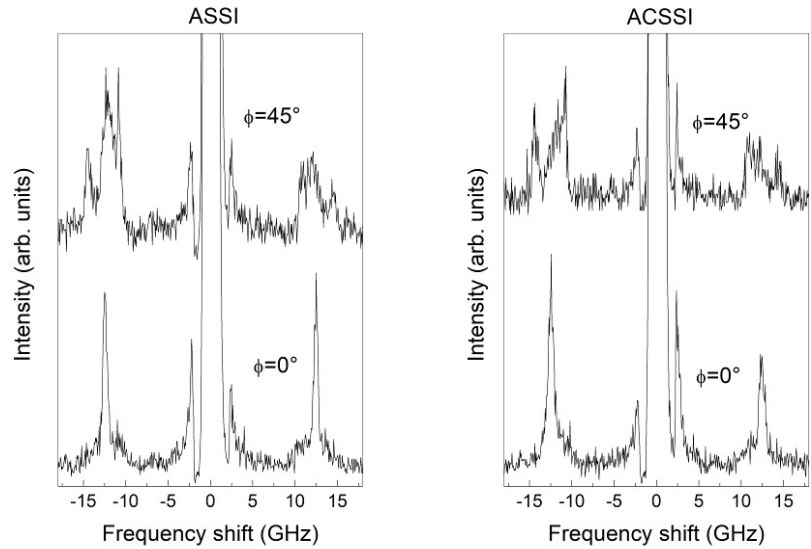

**Fig. 2.** BLS spectra for the ASSI structure (a) and ACSSI structure (b) for the $k = 0$ and the external magnetic applied $H = 0.14$ T oriented at $\phi = 0°$ and 45°.

4. **Theoretical model**

We use the classical continuous medium approach where the magnetization dynamics $\vec{M}$ can be described by the damping-free Landau-Lifshitz (LL) equation:

$$\frac{\partial \vec{M}}{\partial t} = -|\gamma|\mu_0 \vec{M} \times \vec{H}_{eff},$$

where $\gamma$ is the gyromagnetic ratio and $\mu_0$ the vacuum permeability. The effective magnetic field $\vec{H}_{eff}$ consists of the exchange field, the magnetostatic field, and the external magnetic field. Additionally, the magnetic uniaxial anisotropy field is taken into account for the purpose explained below. Assuming the external field to be strong enough to saturate the magnetization of the sample we use the linear approximation thus $\vec{M} = [m_x, m_y, M_S]$, $M_S$ being saturation magnetization while $m_x$ and $m_y$ are the two components of the dynamic magnetization. All vectors are expressed in the Cartesian coordinate system with the z-axis oriented along the external field, the x-axis lying in the plane of the ASI, and the y one pointing out-of-plane direction.

We solve the LL equation by means of the PWM. Since the usage of PWM in magnonics is already elucidated in several papers (see, e.g., Refs. 36,37,38,39) here we recall only the main features of the method. This approach is based on the two transformations which require ideal periodicity and allow to consider only the unit cell with the periodic boundary conditions. Bloch's theorem applies to the dynamic functions, such as the dynamic demagnetizing field and the dynamic components of the magnetization (the system is assumed to be infinite and periodic in the plane of the Py film). The material parameters, namely the saturation magnetization $M_S$, the exchange stiffness constant $A$, and the anisotropy field $H_a$, all being periodic in the real space, are transformed into the reciprocal space by means of the Fourier transform. The solution of the final set of equations is equivalent to the solution of the algebraic eigenvalue problem with the following $2N \times 2N$ block matrix:

$$\widehat{M} = \begin{bmatrix} \widehat{M}_{xx} & \widehat{M}_{xy} \\ \widehat{M}_{yx} & \widehat{M}_{yy} \end{bmatrix}, \qquad (1)$$

where $N$ is the number of plane waves used in the Fourier/Bloch expansion. For the planar MC and the external field aligned in the plane of periodicity the elements of the above matrix are following:

$$M_{ij}^{yy} = -M_{ij}^{xx} = i\frac{1}{H}\frac{k_y + G_{y,j}}{|\vec{k} + \vec{G}_j|} S(\vec{k} + \vec{G}_j, x) M_S(\vec{G}_i - \vec{G}_j)$$

$$M_{ij}^{xy} = M_{ij}^{\Sigma} + \frac{1}{H}\frac{(k_y + G_{y,j})^2}{|\vec{k} + \vec{G}_j|^2}(1 - C(\vec{k} - \vec{G}_j, x)) M_S(\vec{G}_i - \vec{G}_j) \quad (2)$$

$$M_{ij}^{yx} = -M_{ij}^{\Sigma} - \frac{1}{H}C(\vec{k} - \vec{G}_j, x) M_S(\vec{G}_i - \vec{G}_j)$$

where following symbols are used:

$$S(\vec{k},x) = \sinh(kx)\exp(-kd/2),$$
$$C(\vec{k},x) = \cosh(kx)\exp(-kd/2),$$
$$M_{ij}^{\Sigma} = \delta_{ij} + \frac{1}{H}\sum_{l}(\vec{k}+\vec{G}_j)\cdot(\vec{k}+\vec{G}_l)M_S(\vec{G}_i-\vec{G}_l)\lambda_{ex}^2(\vec{G}_l-\vec{G}_j) - \frac{1}{H}\frac{(G_{z,i}-G_{z,j})^2}{|\vec{G}_i+\vec{G}_j|^2}M_S(\vec{G}_i-\vec{G}_j)(1-C(\vec{G}_i-\vec{G}_j,x)).$$

The subscripts *i*, *j* and *l* sweep integers from 0 to N, $\vec{k}$ is the Bloch wave vector of the SW, the vectors $\vec{G}$ are reciprocal lattice vectors, and *H* is the value of the external magnetic field. Here the exchange length is defined as $\lambda_{ex} = \sqrt{2A/\mu_0 M_S^2}$ (an alternative definition is discussed in Ref. 40). Formulas (2) are obtained for the excitations uniform across the film thickness which is a reasonable assumption in our case (the thickness of ASI samples is 20 nm).

Numerical diagonalization of the matrix (1) gives reduced frequencies (eigenvalues) $\Omega = i\omega/|\gamma|\mu_0 H$ and eigenvectors $\vec{m}_{\vec{k}}(\vec{G})$. The last are coefficients of the Bloch expansion of the dynamic magnetization component:

$$\vec{m}(\vec{r}) = \sum_{\vec{G}} \vec{m}_{\vec{k}}(\vec{G})\exp(i(\vec{k}+\vec{G})\cdot\vec{r}).$$

This equation lets us calculate the SW profile, i.e. the spatial distribution of the dynamic magnetization for a given mode and $\vec{k}$. Please notice that both, the *x* and *y*, components of the dynamic magnetization are complex numbers with a phase shift π/2 between the real and imaginary part which gives *T*/4 shift in time, where *T* = 2π/ω is a period of oscillations for a given mode. Usually, distributions of the two components obtained for the same mode are similar thus, it is sufficient to provide just one component (in-plane or out-of-plane) to explain the character of the mode. The eigenvectors allow us to estimate the BLS intensity of the particular modes as well. In the first approximation, BLS intensity is proportional to the square modulus of the fundamental harmonics, i.e., the zero reciprocal vector Bloch coefficient, of the dynamic magnetization $\vec{m}_{\vec{k}}(\vec{G}=0)$.[41]

The important point is that the PWM works properly for bi-component MCs, i.e., when composite consists of two magnetic materials. In the case of ASI there is no magnetic material inside the holes. This leads the LL equation to be undefined in the area of antidots and the method breaks down. The way to avoid this problem is the usage of an artificial magnetic material to be included in the simulation area of the holes. Setting the saturation magnetization of this material much smaller than in the normal magnetic material which is the base of the ASI, splits the solutions into two groups. The first group consists of modes with the dynamic magnetization concentrated in the normal magnetic material and these are physical solutions. On the other hand, the modes in the second group are modes existing mostly in the antidots, thus being unphysical solutions. To eliminate such spurious modes from the spectra, one can shift these unphysical modes up to high frequencies by introducing strong uniaxial magnetic anisotropy inside antidots.[42,43]

## 5. BLS results – experiment and theory

In our calculations, we use material parameters for Permalloy estimated from comparison of calculated dispersion relation with the BLS measurements. The best fit to the experimental data are obtained for the saturation magnetization $M_S = 0.85 \times 10^6$ A/m and the exchange stiffness constant $A = 1.0 \times 10^{-11}$ J/m. For the artificial magnetic material simulating antidots (see the comment above) we set $M_{SA} = M_S / 10^3$, $A_A = A / 10^3$ and the magnetic anisotropy field 50 T (there is no anisotropy field in Py). An external field $H = 0.14$ T is applied along the crystallographic axis of the square lattice of the ASI ($\phi = 0°$) or rotated from this direction by $\phi = 45°$. Since equations (2) stand for the fixed direction of the external field, for our convenience, in the second case the ASI is rotated by 45°, instead of rotating magnetic field. We remark, that the SW propagation perpendicular to the field direction is considered.

In Figs. 3a and b, we plot the comparison between the measured (blue diamonds) and calculated (brown lines) SW frequency dispersion as a function of the wave vector ($k$) for the ASSI array for the two $H$ field orientations ($\phi = 0°$ and $\phi = 45°$). In the theoretical dispersion, we plot only curves with the largest calculated BLS intensity. The structure is based on a square lattice with two antidots in the elementary cell and the lattice constant $a = 900$ nm, so the first Brillouin zone (FBZ) is also square and $0.70 \times 10^7$ rad/m wide. Thus, in the direction along the crystallographic axis ($\phi = 0°$) the edge of the FBZ stands for the wave vector length $k = 0.35 \times 10^7$ rad/m (point X at $k = \pi/a$), while in the direction rotated by $\phi = 45°$ for $k = 0.49 \times 10^7$ rad/m (the distance from the center to the corner M of the FBZ, see top-left insets in Figs. 3a and b). The overall features of the measured dispersions are well reproduced by the PWM calculations: for $\phi = 0°$ (Fig. 3a) we see a set of discrete modes with very small frequency variation in the wave vector range investigated, thus forming well separated flat bands; for $\phi = 45°$ (Fig. 3b) the lowest frequency mode detected in the BLS spectra is characterized by a significant dispersion with a positive group velocity and thus recalling the dispersion characteristic for the DE mode of the continuous (unpatterned) film. However, even in this configuration, two flat bands joining dispersive mode at higher frequency are visible.

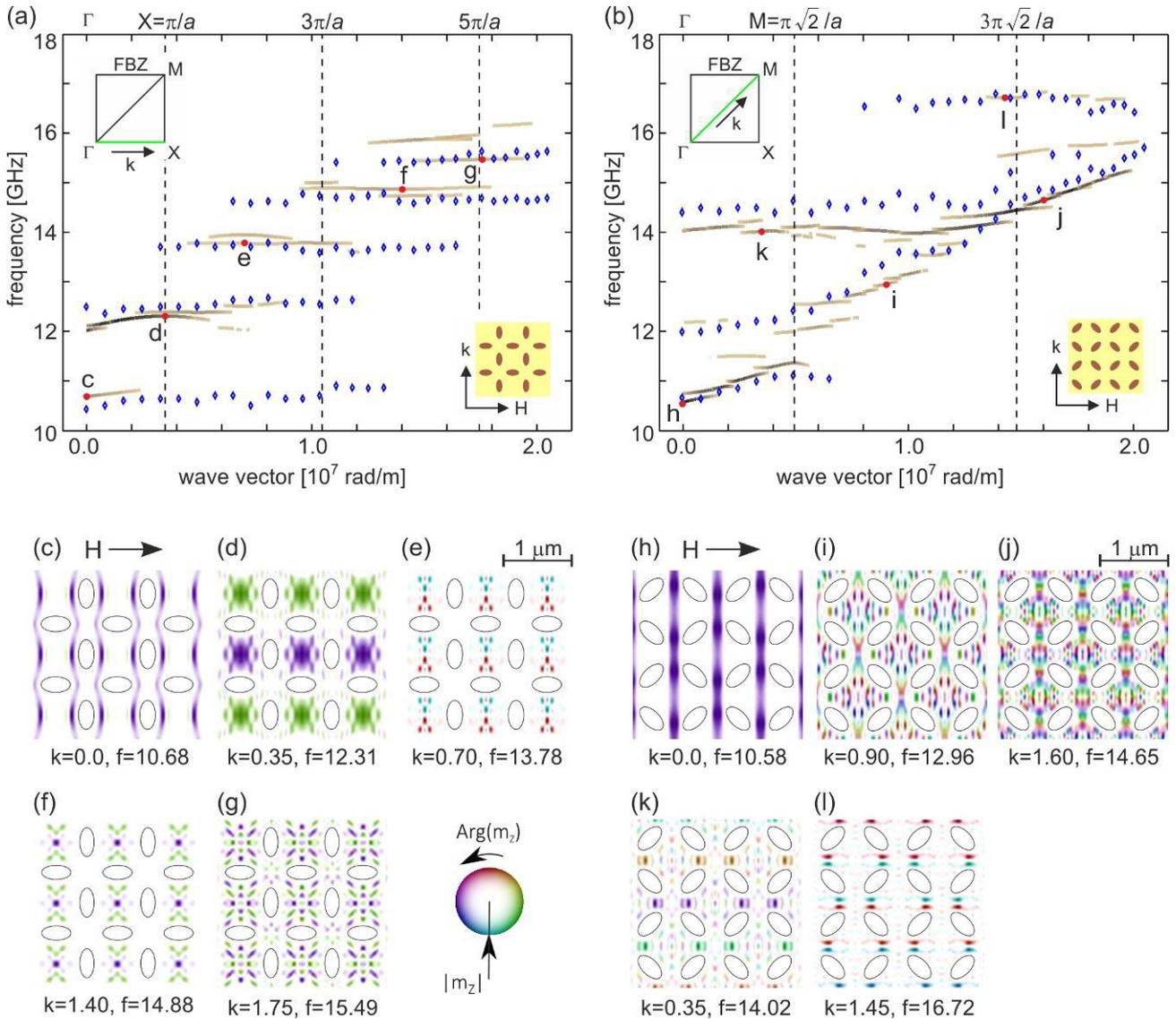

**Fig. 3.** (a, b) Frequency dispersion for the ASSI structure along the direction of a *k*-vector shown in the insets for (a) ϕ = 0° and (b) ϕ = 45°, where ϕ is the angle between the external field *H* and the crystallographic axis of the ASI lattice. Brown lines mark BLS intensity calculated by means of PWM (darker color means higher intensity) and blue diamonds represent experimental data. Borders of the successive Brillouin zones are drawn as vertical dashed lines. (c-l) The profiles of modes marked with red spots in (a) and (b). Each spot is labeled with a letter used to label corresponding SW profile. Colors represent argument (phase) and their intensity the modulus of the dynamic magnetization, as it is shown in the inset. Below each profile, its wave vector *k* (in $10^7$ rad/m) and frequency *f* (in GHz) are specified.

To explain the properties of the measured SW dispersion, first we refer to the spatial distribution of the demagnetizing field, shown in Fig. 4 for the two in-plane directions of the applied magnetic field, and then to the dynamic magnetization spatial profiles shown in Figs. 3c-l. First of all, the Py film is divided by the presence of antidots into two different types of regions marked in Figs. 4 as A and B. The first area (A) has a squared form with four elliptical holes oriented along sides of the square. The second sample portion (B) has a smaller area and it is situated between the edges of four elliptical holes. Another interesting difference is the variation of the demagnetizing field near the edges of antidots in directions along and perpendicular to the external field (this is similar for both field configurations $\phi = 0°$ and 45°, compare Figs. 4a and b). Along the external field, the demagnetizing one continuously changes from zero to minimum (negative, -255 µT) near the border of the hole. This leads to fast but continuous decrease of the effective field. In the perpendicular direction, which is the direction of the SW propagation considered in this paper, the demagnetizing field rises slightly (up to 85 µT) thus the effective field forms sharp edges at antidots borders.

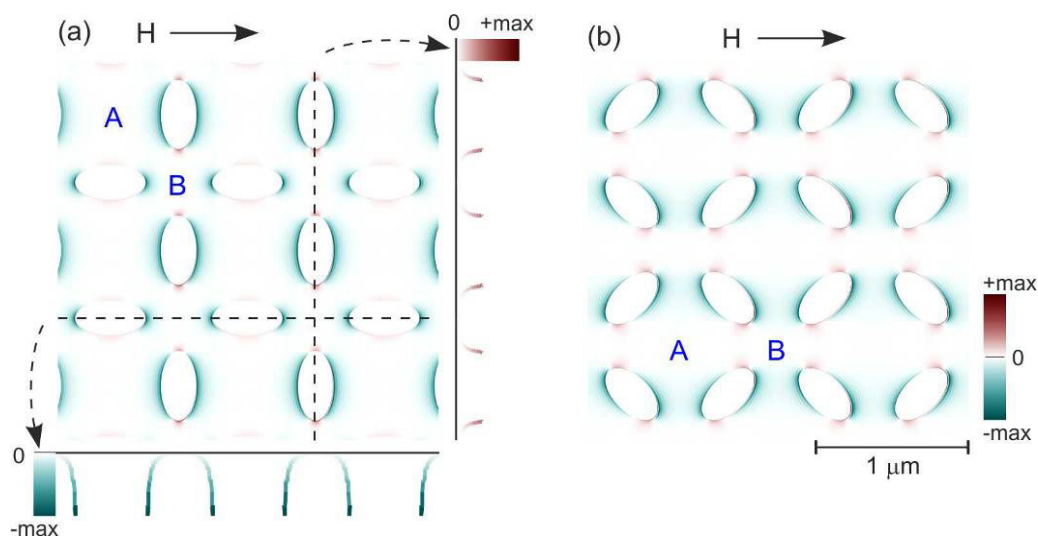

**Fig. 4.** Demagnetizing field for the ASSI structure for (a) $\phi = 0°$ and (b) $\phi = 45°$, where $\phi$ is the angle between the external field *H* and the crystallographic axis of the ASI lattice. In the Py area, the two regions marked with letters A and B are distinguished.

In Figs. 3c-g we show the SW profiles of some selected modes for the ASSI structure for $\phi = 0°$ (the modes are marked in Fig. 3a with red spots and labeled with respective letters). The dynamic magnetization calculated by means of the PWM is complex number so in all profiles the color of the plot refers to the argument (phase) of the magnetization and its intensity to the modulus (the amplitude of the magnetization precession) as it is depicted in the inset – darker color means bigger amplitude while white negligibly small. Each mode is a typical and highly intense representative of one of the BLS bands. The first profile is taken for the lowest mode visible in the BLS spectrum for *k* = 0 (Fig. 3c). The magnetization precession is all-in-phase along the direction of k and has a large calculated BLS intensity up to $k$ about 0.23 x $10^7$ rad/m (Fig. 3a). The amplitude of this mode is confined to small areas along sides of antidots orthogonal to the field direction and strongly concentrated in regions A (compare Fig. 4a). Here we would like to notice that terms 'concentration' and 'localization' are not equivalent. Localization is accompanied with the exponential decay of the amplitude of the magnetization precession (at least in some directions), it usually refers to localization on defects (point, line or plane defects). Concentration depicts the tendency of the dynamic magnetization just to enhance the amplitude in some areas rather than the others, which can be replicated in each unit cell. Of course, localization entails concentration, usually very strong, but concentration does not necessarily mean localization. Next profiles shown in Figs. 3d-g are taken for the higher BLS bands. All these modes are confined almost solely to regions A. In general, the profiles of these modes have SW amplitude concentrated in the regions between antidots. As a consequence, an individual SW has very small group velocity and the bands are flat in the BLS spectra, seen in Fig. 3a.

In Figs. 3h-l the SW profiles of some particular modes for ASSI structure with 45° rotated external magnetic field are shown (depicted in Fig. 3b with red spots and corresponding letters). Again, the first one (panel h) is the lowest mode visible in BLS but now the profile shows the amplitude concentrated in well-defined channels confined between edges of the antidots.[21,44] Please notice that the profile in Fig. 3h is taken for the FBZ center, i.e. for the wave of infinite wavelength. If we move to $k > 0$ the mode becomes propagating but its profile is still concentrated in channels along the wave vector. Within these propagation channels, about 240 nm wide, the group velocity estimated from PWM is about 610 m/s, which results in a broad magnonic band. This leads to a series of anti-crossings with higher modes with the typical curvature of the frequency branches found in calculated spectra in Fig. 3b. The influence of some 'soft' constraints could be seen as a non-uniformity of the profile along the direction of propagation. The concentration of the amplitude is stronger in the regions between two antidot edges laying close to each other in the direction of the external field (horizontally). This feature is due to the demagnetizing field near the antidot edges. As one can see in Fig. 4b at these close-together edges the demagnetizing field is negative which causes the lowering of the effective field and finally attract the SW excitation. Both features, the propagation within channels and the concentration near antidot edges, are preserved when passing through anti-crossings as it is shown in Figs. 3i and j, where profiles of two exemplary higher modes are shown (the group velocities estimated from PWM results are about 500 m/s).

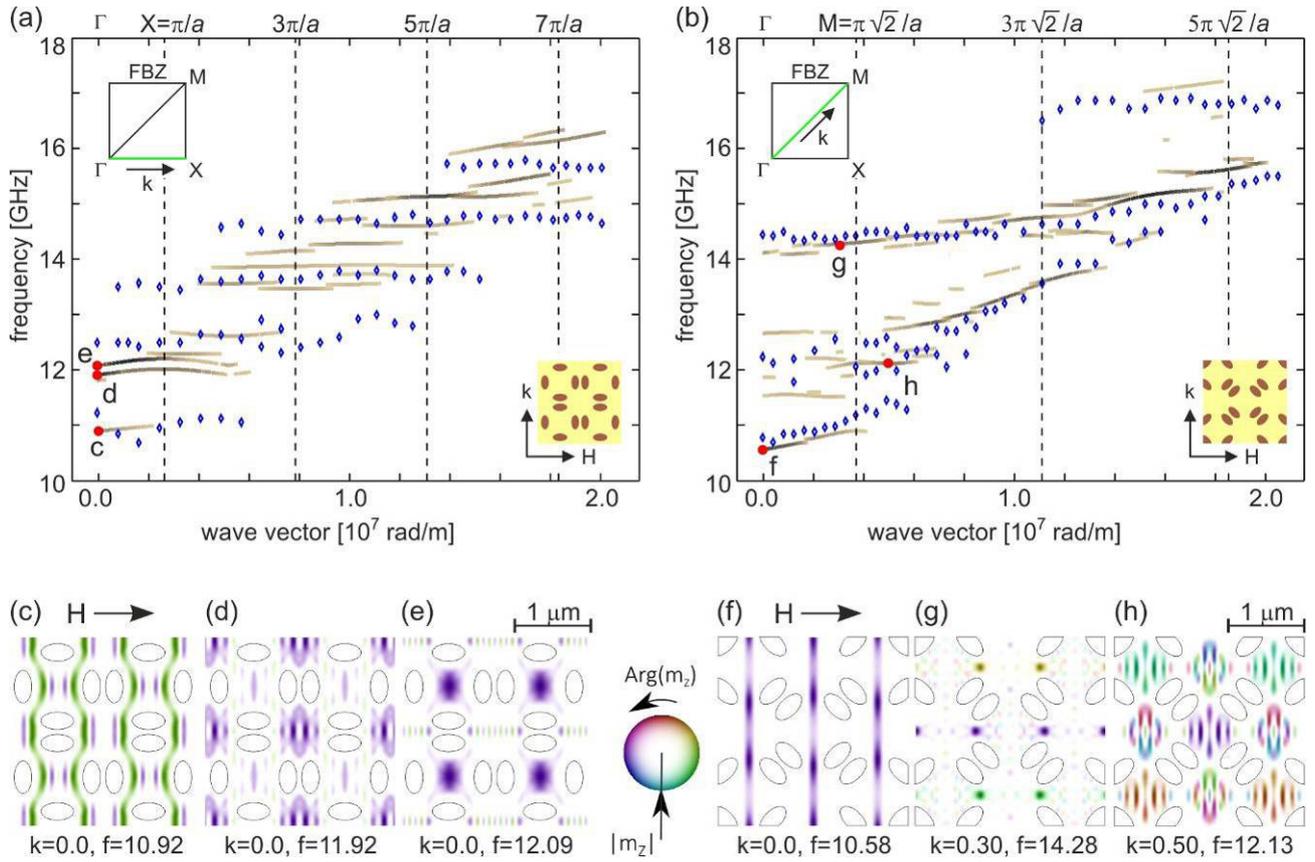

**Fig. 5.** (a, b) BLS spectra for the ACSSI structure along the direction of a *k*-vector shown in insets for (a) ϕ = 0° and (b) ϕ = 45°, where ϕ is the angle between the external field *H* and the crystallographic axis of the ASI lattice. Brown lines mark BLS intensity calculated by means of PWM (darker color means higher intensity) and blue diamonds represent experimental data. Borders of successive Brillouin zones are drawn as dashed lines. (c-h) The profiles of modes marked with red spots in (a) and (b). Each spot is labeled with a letter used to label corresponding SW profile. Colors represent argument (phase) and their intensity the modulus of the dynamic magnetization, as it is shown in the inset. Below each profile its wave vector *k* (in $10^7$ rad/m) and frequency *f* (in GHz) are specified.

The shape of the demagnetizing field determines also the existence of two flat bands in the SW spectrum around 14.1 and 16.7 GHz for the tilted external field (Fig. 3b). The SW profiles for these bands are shown in Figs. 3k and l. In contrary to the ascending modes, here we have a concentration of the SW amplitude between antidot edges which are close to each other along the direction of the SW propagation (perpendicular to the external field). At these edges the demagnetizing field is positive, and thus in this direction we have very sharp edges of the effective magnetic field (see Fig. 4b). The concentration of the amplitude in the small areas between antidots leads to the similar effects as it was in the case of the non-rotated field: a limited range of frequencies available for an individual SW. So again, related frequency bands are flat.

For ACSSI structure in Fig. 5a and b we compare the experimental (blue diamonds) and theoretical (brown lines) SW dispersion for both directions of the external magnetic field. In calculated spectra, only high BLS intensity modes are shown (the darker color depicts higher calculated BLS intensity). Successive Brillouin zone borders are marked by dashed lines. The lattice constant equals 1200 nm thus for $\phi = 0°$ the FBZ border is at $k = 0.26 \times 10^7$ rad/m (point X) and for $\phi = 45°$ the FBZ corner is at $k = 0.37 \times 10^7$ rad/m (point M). The results are in good agreement and the overall tendency is the same as it was for ASSI structure – dominating flat bands for the external field along the main crystallographic axis and ascending ones for the rotated field. The difference is that now we have more modes visible in BLS spectra. Especially in Fig. 5a the calculated frequency branch around 12 GHz is doubled (compare Fig. 3a) and in Fig. 5b in the range of frequencies between 11.5 and 13.0 GHz experimental points, as well as calculated ones, are scattered rather than arranged in bands.

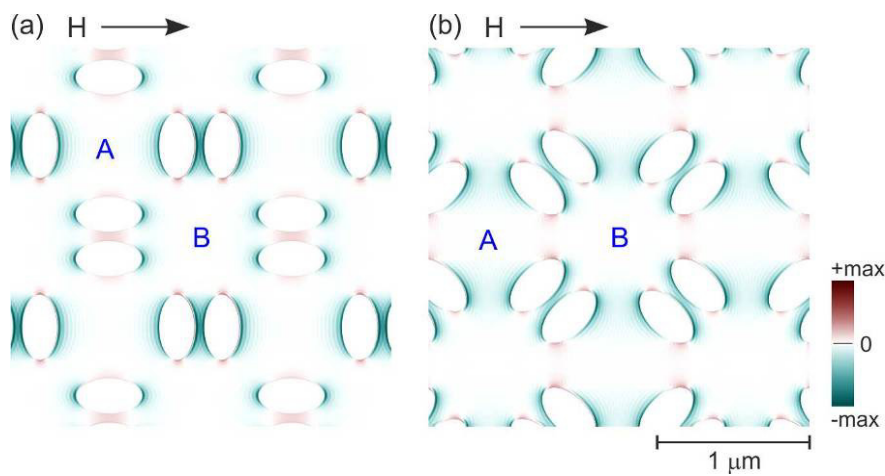

**Fig. 6.** Demagnetizing field for the ACSSI structure for (a) $\phi = 0°$ and (b) $\phi = 45°$. The Py area is divided into two different regions marked with letters A and B.

Main effects are like it was in the case of ASSI because the changes in the spectra are consequences of the change of shape of the demagnetizing field, which is similar for both structures (compare Figs. 4 and 6). Again, for the unrotated external field the potential wells for SWs created by this field are formed between antidots while for the field rotated by 45° the continuous channels appear along the direction of the SW propagation (perpendicular to the field).

The influence of the demagnetizing field on the SW profiles for ACSSI is demonstrated also in Figs. 5c-h for six representative modes marked in Fig. 5a-b with red spots and respective letters. Profiles c, d, and e are for the unrotated external field (all of them for $k = 0$). The first profile stands for the lowest mode visible in BLS. In comparison with the similar mode in ASSI (Fig. 3c) neighboring maxima of the dynamic magnetization are closer together (four maxima along 2.5 μm in the vertical direction for ACSSI while three for ASSI) which results in a bit higher frequency. Next two modes are concentrated in regions B and A, respectively. In ACSSI structure both areas A and B are similar which results in the occurrence of pairs of modes with similar frequencies and BLS intensities – one concentrated in the region A and next in the region B. For ASSI the areas A and B were very different and such twin-modes did not exist.

For $\phi = 45°$ the lowest mode in the BLS spectrum (Fig. 5f) is analogous to that in ASSI, i.e. freely propagating in the channels between antidots (with the group velocity estimated from PWM results c.a. 520 m/s). Consequently, the lowest visible mode ascends in frequency with increasing wave vector. Alike in ASSI, for the ACSII structure also localized modes exist. An exemplary profile is shown in Fig. 5g for a mode forming a flat band just above 14 GHz in the spectrum in Fig. 5b. The very rich BLS spectrum between 11.5 and 13.0 GHz is a result of similar areas of Py in regions A and B in ACSII. As we have already noticed, this causes additional modes to be of high BLS intensity – see twin modes d and e in Fig. 5. An additional example is shown in Fig. 5h. This profile is spread out in both regions thus exhibits a rather non-propagative character.

## 6. Tailoring of the frequency spectrum

To gain more insight into the SW spectra in ASSI and ACSSI, we perform a detailed analysis of the calculated frequency dispersion of all modes. Full information about the magnonic band spectra will be useful for further investigations of the SW transmission in ASSI or ACSSI, where the multi-modal interaction could take place and influence the propagation of waves. The results of calculations for both directions of the external field $H$ are shown in Fig. 7 (a) and (c) for $\phi = 0°$ and in Fig. 7 (b) and (d) for $\phi = 45°$, for both samples: ASSI (panels a and b) and ACSSI (panels c and d). All spectra are limited to the FBZ and up to 13 GHz, only. The color of lines depicts calculated BLS intensity: olive branches will be invisible in BLS while brown shall be visible – darker line means greater intensity. The additional complexity of the magnonic band structures is present here which was not visible in the BLS spectra in Figs. 3 and 5, where only lines with high intensity were shown. The SW band structures are very dense in terms of the number of magnon states per unit frequency, but still, directional magnonic band gaps are visible (blue rectangles). Interestingly, for $\phi = 0°$, the band gaps are opened only at the Brillouin zone boundary or center, while for $\phi = 45°$ they appear also at intermediate *k*-values, being the widest (ASSI) and the second widest (ACSSI) gaps in the spectra (see Fig. 8 for an explanation). As it was shown in the previous section for ϕ = 0° case, SW profiles are concentrated either in region A or B. In ASSI these areas are very different while in the ACSSI they are similar (see Figs. 4 and 6), thus it leads to the opening of new band gaps in the ACSSI (compare Figs. 7a and c), and separation of both types of excitations (confined to regions A or B). For ϕ = 45° most of these gaps are closed (Fig. 7d) due to the propagation of SWs in the channels, discussed already in the previous section.

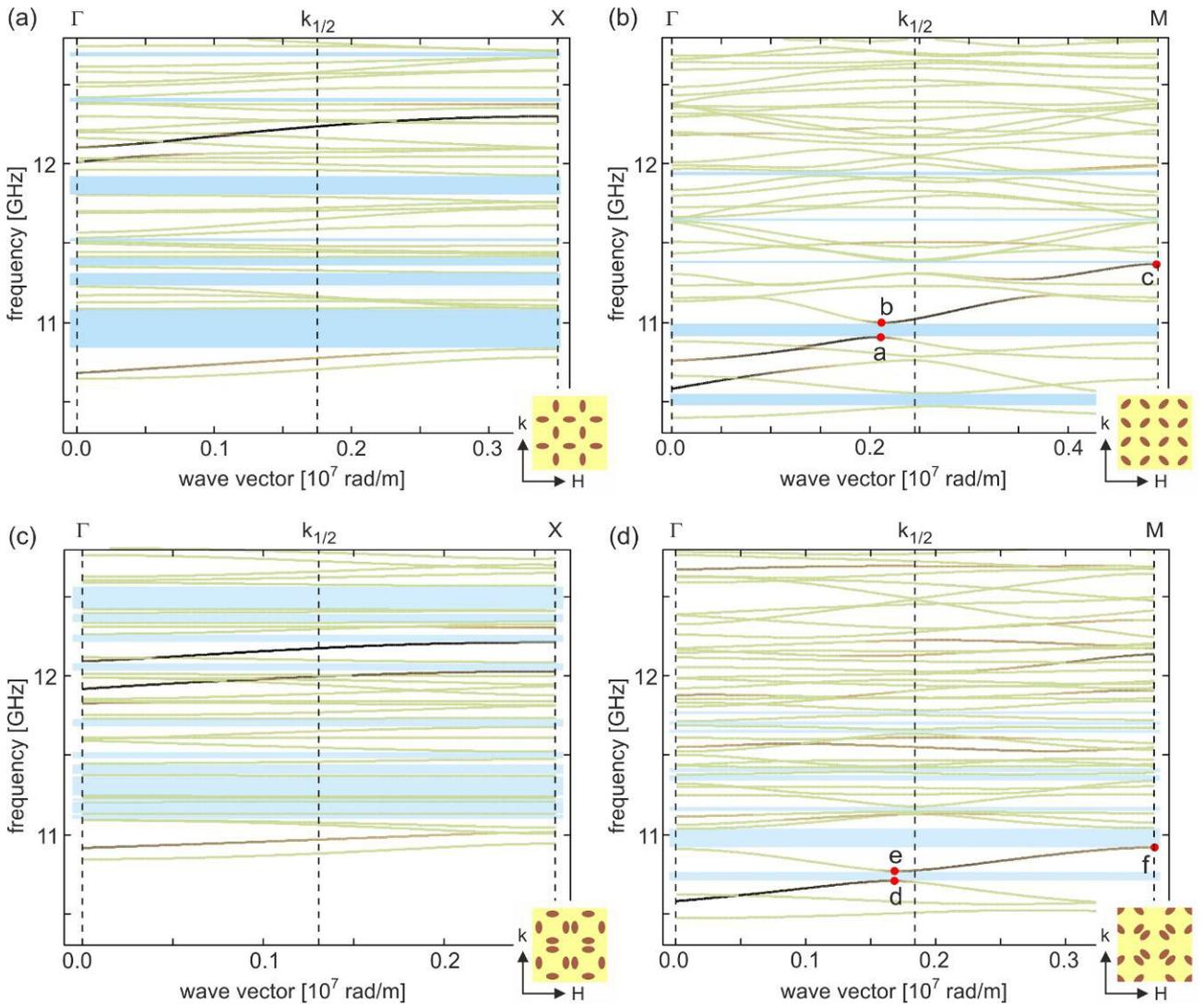

**Fig. 7.** PWM results of the SW spectrum within the FBZ (a, b) for ASSI and (c, d) ACSSI structures along the direction of the *k*-vector shown in the insets for (a, c) ϕ = 0° and (b, d) ϕ = 45°. The color intensity is related to the calculated BLS intensity, brown lines mark high intensity while olive branches low. Blue bars represent magnonic band gaps. Dashed lines stand for the center (Γ), the middle ($k_{\frac{1}{2}}$), and the border (X, M) of the FBZ. Red spots in the spectra (b) and (d) labeled with letters mark the modes, the profiles of which are shown in Fig. 8.

In Figs. 8a and b we show profiles of modes delimiting the gap at $k$ = 0.21 x $10^7$ rad/m (ASSI, see red spots in Fig. 7b). These two profiles have similar spatial distribution, but they are excited in different regions. Such property is typical for the Bragg band gap opening at the Brillouin zone border, where both modes delimiting the gap should have profiles concentrated mostly in one region only, A or B, thus for the upper edge of the gap the amplitude is shifted by a half of the period, which is consistent with the approximate models.[45,46] This suggests, that the periodicity of the structure for this SW is doubled. The effect is a consequence of two similar antidots in an elementary cell. Both of them are ellipses of the same shape, but they are rotated by 90° with respect to each other, so they are not equivalent. As a result, the exact periodicity in the direction of the SW propagation at 45° appears every second line of antidots, while two neighboring lines are shifted alongside.

On the other hand, the neighboring lines of holes are very similar which interfere with the doubled periodicity. As a consequence, the phase shift between two neighboring high amplitude lines is not exactly $\pi/2$ but only $0.43\pi$. Taking into account just single propagation channel adjacent maxima differ in phase by $0.86\pi$ but their spatial separation is doubled in comparison to the original square lattice of ASSI structure. This is compatible with the wave vector the extra band gap in Fig. 7b appears for. This vector is not a half of this at the FBZ border ($k_{\frac{1}{2}} = \sqrt{2}\pi/2a$ for $\phi = 45°$), as it should be in the case of doubled periodicity, but it is 86% of $k_{\frac{1}{2}}$. In addition, the lowest band gap in this spectrum has a similar origin and appears for the wave vector bigger than $k_{\frac{1}{2}}$. In fact, several modes in Fig. 7b look like plotted over more than one Brillouin zone, but they are not strictly symmetric and the 'intermediate point' is not at $k_{\frac{1}{2}}$. We attribute this lack of the full symmetry also to the rotation of adjacent antidots.

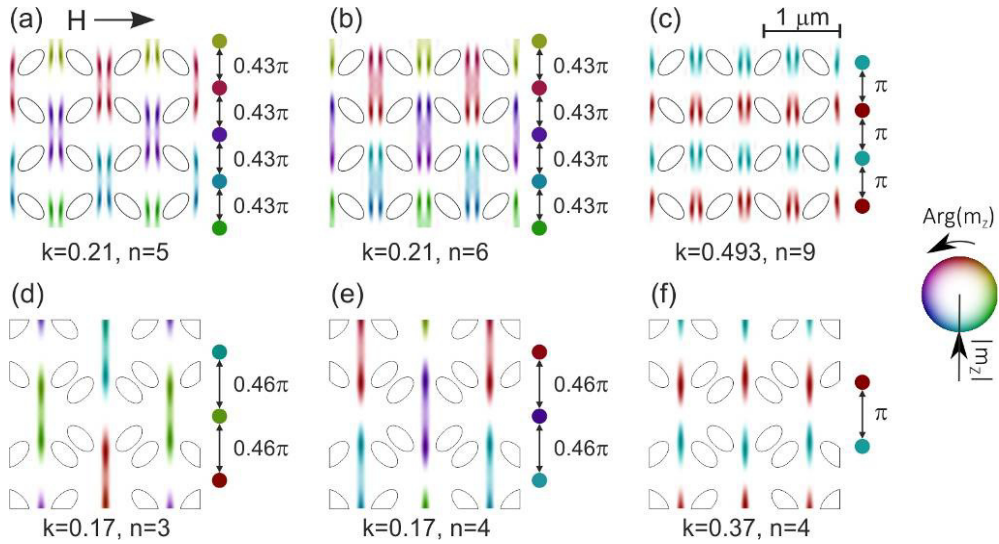

**Fig. 8.** SW profiles of the modes bounding the lowest magnonic band gaps for $\phi = 45°$ in (a-c) ASSI and (d-f) ACSSI structures. The bounds of the gaps are marked with red spots and respective letters in Fig. 7. Colors represent argument (phase of SW) and their intensity modulus of the dynamic magnetization, as it is shown in the inset. The change of the phase along the vertical direction within each profile is also provided on the right side of each plot. Below each profile its wave vector $k$ (in $10^7$ rad/m) and the mode number in the frequency spectrum $n$ are specified.

Obviously, the original periodicity of the square lattice is maintained for any propagation direction. The SW profile at the border of the FBZ (for 45°-propagation) of the original square lattice of ASSI structure is shown in Fig. 8c. This is the mode from the bottom bound of a Bragg gap. Typically for the square lattice, the distance between neighboring lattice planes is $a/\sqrt{2}$, which gives the separation of neighboring rows of antidots. According to the profile in question, the phase shift of the dynamic magnetization at this distance is $\pi$, as it should be.

Figs. 8d-f present three modes delimiting band gaps in the case of rotated field in ACSSI. Modes d and e are lower and upper bound of the gap opening around the middle of the FBZ. Their properties are similar as it was for ASSI structure – they are concentrated in different regions. However, the band gap opening is at 92% of $k_{\frac{1}{2}}$ (see Fig. 7d) and their phase changes by $0.92\pi$, if the space distance doubles with respect to the original square lattice of ACSSI. Again, we relate this to the shape of the antidots which contains a couple of elliptical holes for ACSSI and there is a rather small difference between such double ellipses rotated left or right. This difference is smaller than for an ASSI, which results also in a narrower gap. It suggests that playing with the similarity of neighboring lines (e.g. by rotation of antidots) one can tune the width of the gap and the wave vector at which it opens.

The last profile in Fig. 8, f, is for the lower bound of the next band gap (the widest in the spectrum) at the FBZ border and it has appearance typical for the Bragg gap with the opposite phase at neighboring crystallographic lines perpendicular to the propagation direction.

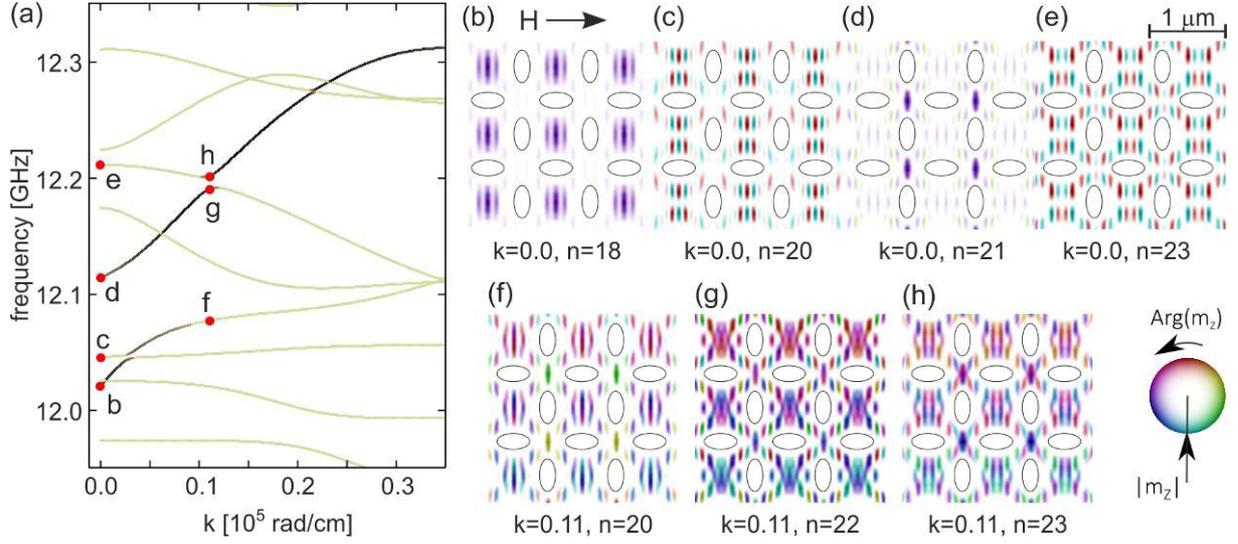

**Fig. 9.** SW profiles for the ASSI structure for $\phi = 0°$. (a) Enlarged part of the spectrum form Fig. 7a between 11.95 and 12.35 GHz. (b-h) Profiles of hybridizing modes marked with spots in (a). Each spot is labeled with a letter used to label corresponding SW profile. Colors represent argument (phase) and their intensity the modulus of the dynamic magnetization, as it is shown in the inset, and the mode number in the frequency spectrum is given by $n$.

Another interesting effect appearing in ASSI structure for the unrotated field is a multimode hybridization. The group of hybridizing modes lies between 12.0 and 12.3 GHz in the spectrum shown in Fig. 7a. This part of the spectrum is enlarged in Fig. 9a, where the spots and the corresponding letters stand for the respective profiles shown in Figs. 9b-h. Among the others, there is a fundamental mode in this group (b), which is a counterpart of a uniform excitation having dynamic magnetization all in phase. It starts slightly above 12 GHz for $k = 0$ as a mode with number $n = 18$ and the profile is shown in Fig. 9b. On increasing the frequency of the fundamental mode first crosses mode $n = 19$ and then hybridizes with the mode $n = 20$ (Fig. 9c). The effect of this interaction is clearly seen in Fig. 9f, where we provide the profile of the $n = 20^{th}$ mode – in the region A the profile is a mixture of profiles b and c. Detailed inspection of the figure shows, that this SW mode is also mixed with some higher modes (pointed by a non-zero amplitude in the region B) due to the next hybridization, where three modes are interacting simultaneously. (Similar effect, namely three mode hybridization and the matching of hybridizing modes profiles can be found also in other systems, e.g., in circularly magnetized disks and rings.)[47,48] Among these three modes, there is one which starts at $n = 21$ and $k = 0$ with the profile strongly concentrated in region B (Fig. 9d). Please notice that this mode is not of fundamental kind, because there are small peaks with opposite phase near the corners of the region A. This mode passes through the spectra with a positive slope having strong and almost constant BLS intensity. With growing wave vector after crossing next higher mode, it hybridizes with the mode *n* = 23 (the profile for *k* = 0 is shown in the panel e) around *k* = 0.11 x $10^7$ rad/m (Figs. 9g and h). In this hybridization, the fundamental mode is involved, as can be seen in Fig. 9f – the strong dynamic magnetization amplitude in the region B indicates the influence of the profile d. This is also reflected in the anti-crossing of the modes in question with its characteristic curvature of frequency branches.

## 7. Conclusions

In the paper, we have investigated experimentally with BLS, and theoretically with PWM the dispersion relation of SWs propagating perpendicular to the external magnetic field in two anti-ASI structures consisting of single (ASSI) and double (ACSSI) elliptical holes in Py film arranged into a squared lattice. Overall, good agreement between calculated and measured dispersion relation has been found. We demonstrated, that by rotation of the external magnetic field by 45° we can transform the magnonic spectra from consisting of predominantly non-propagative waves into the spectra of propagating SWs with a number of hybridizations. We showed, that in both structures the Py film with antidots can be regarded as split into small regions A and B, surrounded by long and short sides of the elliptical holes, respectively. These regions can be treated as isolated or connected depending on the SW propagation direction. This results in three main types of the SW profiles. For propagation along the main axis of the antidot lattice the SWs are strongly confined to the regions A and B, which leads to flat bands and several band gaps in the spectrum. For 45° direction the opening of propagation channels for SWs leads to broadening of the bands consisting of modes excited along these channels. In this case, the majority of band gaps are closed. The third type of modes has the profile mostly localized at the ends of neighboring antidots close to the deep wells of the demagnetizing fields. Strongly constrained profiles cause only slight changes of the frequency with the wave vector. Thus, the specific structure of the ASI leads to the channeling of the SW propagation in certain directions.

The specific arrangement of antidots in an elementary cell results in doubled periodicity for 45° propagation direction. This leads to the additional band gaps opening close to the middle of the FBZ ( $k_{\frac{1}{2}} = \sqrt{2}\pi/2a$ ). We attribute the shift of the band gap bounds from $k_{\frac{1}{2}}$ to the difference between neighboring antidots, which are rotated by 90°. This difference is smaller in the case of the structure with double ellipses (ACSSI), which leads to a smaller shift and narrower gaps. Of course, the original periodicity of the square lattice is retained, thus the Bragg band gap occurs at the FBZ border with the specific SW profile of modes at its bounds. For ACSSI, based on pairs of antidots instead of single ones, playing with the distance between antidots in a pair should be useful way for the tuning of the band gaps in the SW spectrum via smooth change of the area of region B. These properties demonstrate the way for a fine tuning of the SW spectra in planar magnonic structures.

All the above features originate in the specific structure of the ASI systems in which the lattice of antidots divides the thin film into regions. Such regions are not present in the standard antidot lattices.

**Acknowledgments**

The study has received financial support from the National Science Centre of Poland under grant No. UMO-2016/21/B/ST3/00452 and the EU's Horizon 2020 research and innovation programme under Marie Sklodowska-Curie GA No. 644348 (MagIC). Grant funding from the Ministry of Education Singapore, under research Project No. R-263-000-C61-112, is gratefully acknowledged. A.O.A. is a member of Singapore Spintronics Consortium (SG-SPIN).